\documentclass[12pt]{article}
\usepackage{color}
\usepackage{amsmath, amssymb}
\usepackage{graphics}

\begin{document}
\baselineskip=16pt

\begin{center}

{\Large\bf Baryogenesis from Dark Sector}

\vspace{1.2cm}

{\large N. Haba}$^1$
and 
{\large S. Matsumoto}$^{2,3}$

\vspace{0.7cm}

$^1${\it Department of Physics,
         Osaka University, Toyonaka,
         Osaka 560-0043, Japan} \\

$^2${\it Department of Physics,
         University of Toyama,  
         Toyama, 930-8555, Japan}\\

$^3${\it IPMU,
         The University of Tokyo,  
         Kashiwa, 277-8583, Japan}\\

\vspace{0.7cm}

{\bf Abstract}

\vspace{0.5cm}

{\parbox{15cm}{\noindent
We propose a novel mechanism to generate a suitable baryon asymmetry from dark (hidden) sector. This is a Baryogenesis through a reverse pathway of the ``asymmetric dark matter'' scenario. In the mechanism, the asymmetry of dark matter is generated at first, and it is partially transferred into a baryon asymmetry in the standard model sector. This mechanism enables us not only to realize the generation of the baryon asymmetry but also to account for the correct amount of dark matter density in the present universe within a simple framework.}}

\end{center}

\vspace{0.8cm}

\noindent
{\large{\bf Introduction}}\\

\vspace{-0.5cm}

\noindent
Thanks to recent cosmological observations, it has been revealed that the energy density of dark matter is five times larger than that of baryon, $\Omega_{DM} \sim 5\Omega_B$~\cite{Komatsu:2010fb}. The ``asymmetric dark matter (ADM)'' scenario~\cite{Kaplan:2009ag} gives one of interesting explanations to the result. In the ADM scenario, the dark matter is assumed to have the mass of ${\cal O}(10)$ GeV and be in chemical equilibrium with SM particles in the early universe. The relation $\Omega_{DM} \sim \Omega_B$ is then naturally obtained because the number densities of dark matter and baryon are in the same order due to the equilibrium. On the other hand, the ADM scenario is simply realized when the anti-dark matter exists (in addition to the dark matter) and there is a messenger interaction between dark (hidden) and standard model (SM) sectors. The interaction plays an important role to maintain the chemical equilibrium. Because of this excellent explanation between dark matter and baryon densities, several studies have been performed so far~\cite{{Cohen:2010kn}, {Hooper:2004dc}, {Frandsen:2010yj}}. A similar scenario has also been proposed in a context of technicolor-like setup~\cite{Kribs:2009fy}.

In the ADM scenario, the baryon (lepton) asymmetry is assumed to be generated at first by an appropriate mechanism and translated into dark matter asymmetry. However, as can be seen in conventional Baryogenesis scenarios, the generation of the asymmetry often has some difficulties due to the existence of experimental constraints. For example, the Leptogenesis scenario, which is one of the most attractive scenarios for the baryon asymmetry of the universe, is severely constrained~\cite{Davidson:2002qv}.

In this letter, we propose a novel mechanism to generate a suitable baryon asymmetry from dark sector~\cite{note added}. This is a Baryogenesis through a reverse pathway of the ADM scenario. At first, the asymmetry of the dark matter is generated in the early universe, and after that, it is transferred into a suitable baryon asymmetry in the SM sector. This mechanism generates not only the baryon asymmetry but also the correct amount of dark matter density. Since the dark sector does not receive severe constraint from current experiments, we can easily construct a model to generate the dark matter asymmetry. This is a contra-distinctive feature to conventional scenarios.
\\

\noindent
{\large{\bf Dark sector}}\\

\vspace{-0.5cm}

\noindent
First, we show our setup in the framework of supersymmetry (SUSY). We introduce $X$, $\bar{X}$ and $Y_i$ ($i = 1, 2$) fields into the next-to-minimal supersymmetric standard model (NMSSM). These fields are singlet under SM gauge groups, and fermionic components of $X$ and $\bar{X}$ correspond to dark and anti-dark matter particles, respectively. Scalar components are expected to be heavier than the fermionic ones due to soft SUSY breaking terms. $Z_{4R}$ symmetry, which is a part of $U(1)_R$, and the lepton number symmetry ($U(1)_L$) are imposed in the Lagrangian, and we postulate that only the $U(1)_L$ symmetry is softly broken. Charge assignments of the fields are as follows.\\
\begin{center}
\begin{tabular}{c|ccc}
\hline
& $X$ & $\bar{X}$ & $Y_i$ \rule{0ex}{2.2ex}  \\
\hline
$Z_{4R}$ & $i$ & $-i$ & $-1$  \\
$U(1)_L$ & $1/2$ & $-1/2$ & $1$  \\
\hline
\end{tabular}
\end{center}

\vspace{0.3cm}

\noindent
With the charge assignments above, the superpotential is written by
\begin{eqnarray}
\mathcal{W} =
\mathcal{W}_{\rm NMSSM}
- \frac{M_i}{2} Y_i Y_i - m X \bar{X}
+ \frac{\kappa_i}{2} Y_i \bar{X}^2
+ \lambda S X \bar{X}
+ \frac{y_i}{2\Lambda} \bar{X}^2 L_i H_u
+ \frac{y_i^\prime}{2\Lambda} \bar{X}^2 Y_i S,
\end{eqnarray}
where $L_i$ is the $i$-th generation ($i=1,2,3$) lepton doublet, $H_u$
is the Higgs doublet giving the masses of up-type quarks, and $S$ is the
singlet field predicted in the NMSSM. The superpotential of the NMSSM is
denoted by $\mathcal{W}_{\rm NMSSM}$. In the superpotential, we write
down operators up to ${\mathcal O}(1/\Lambda)$, where $\Lambda$ is an
energy scale characterizing the strength of interactions that break the
lepton number of dark sector. There exist other operators of this order
which does not break the number. They are, however, not relevant to
following discussions, and we omit writing those operators
explicitly. Mass matrix of $Y_i$ has already been diagonalized and whose
mass eigenvalues $M_i$ as well as the dark matter mass $m$ are real and
positive by appropriate redefinitions of $Y_i$, $X$ fields. In our
setup, we consider a case where $Y$ is much heavier than $X$ and
$\bar{X}$. One of the coupling constants $\kappa_i$ is still complex in this basis, which will be the origin of the dark matter asymmetry. On the other hand, the non-renormalizable interaction $\bar{X}^2 L_i H_u$, which is called the ADM messenger interaction in following discussions, plays a crucial role to mediate dark matter to baryon asymmetries in the early universe.

Here, we estimate the energy scale of $\Lambda$ which enables to mediate the asymmetry. First, the expansion rate of the universe is determined by the Friedman equation. With the Planck scale being $M_{pl} \simeq 10^{19}$ GeV, the Hubble parameter $H$ is given by
\begin{equation}
H \simeq 1.66 \sqrt{g_*} T^2/M_{pl},
\label{Eq:Hubble}
\end{equation}
where $g_* \sim {\mathcal O}(100)$ is the massless degrees of freedom in the early universe. Next, the reaction rate of the process through the ADM messenger interaction is given by
\begin{equation}
\Gamma_{\Lambda} \simeq |y_i|^2 T^3/(8 \pi \Lambda^2). 
\label{Eq:ADMop}
\end{equation}
When the energy scale of $\Lambda$ is small, coupling constants $y_1$ and $y_2$ should be suppressed to evade constraints from lepton flavor violating processes. Finally, the reaction rate of the Sphaleron process exceeds the expansion rate of the universe (the Hubble parameter $H$) when the temperature of the universe is within the range,
\begin{eqnarray}
100~{\rm GeV} \leq T \leq 10^{12}~{\rm GeV}.
\label{Eq:Spharelon}
\end{eqnarray}
With the use of Eqs.(\ref{Eq:Hubble})-(\ref{Eq:Spharelon}), the upper bound on the scale $\Lambda$ is obtained, because both Sphaleron and ADM messenger interactions should be active to transfer the dark matter asymmetry to the baryon asymmetry. We therefore obtain the condition $\Gamma_\Lambda \geq H$, at least, at $T \sim 10^{12}$ GeV. This condition leads to the upper bound on $\Lambda$ as
\begin{eqnarray}
\Lambda \leq 10^{14}~{\rm GeV}, 
\end{eqnarray}
with the coupling constant $y_3$ being ${\cal O}(1)$. The bound indicates that the energy scale of $\Lambda$ must be lower than the GUT scale. On the other hand, if we assume that the ADM messenger interaction decouples before the Sphaleron process becomes inefficient, namely, at $T \leq 100$ GeV, the lower bound on the scale $\Lambda$ is obtained as
\begin{eqnarray}
\Lambda \geq 10^9~{\rm GeV}.
\end{eqnarray}

\noindent
{\large{\bf Darkgenesis}}\\

\vspace{-0.5cm}

\noindent
We are now at the position to discuss the generation of the baryon asymmetry and the dark matter density in the framework of the scenario mentioned above. First, at the very early universe, the lighter mass eigenstate $Y_1$ is in thermal equilibrium. Here, we are assuming the hierarchy between $Y_1$ and $Y_2$ to be $M_1 \ll M_2$ to make the discussion simple.  In addition, the mass of $Y_1$ ($M_1$) is set to be as low as $10^6$ GeV, and assume that the reheating temperature of the universe after inflation can be low enough to avoid dangerous gravitino problem~\cite{Kawasaki:2008qe}. The $Y_1$ particle eventually decays into two $\bar{X}$s, where the asymmetry of the dark matter number ($DM$) is also generated due to $CP$ phases of the coupling $\kappa_i$. The mechanism is the same as the traditional Leptogenesis scenario, and Boltzmann equations to describe this phenomenon are
\begin{eqnarray}
&&
\dot{n}_Y + 3 H n_Y = -\Gamma_D (n_Y - n_Y^{EQ}),
\nonumber \\
&&
\dot{n}_X + 3 H n_X = \epsilon \Gamma_D(n_Y - n_Y^{EQ})
- ({\rm Washout~terms}),
\end{eqnarray}
where $n_Y$ is the summation of number densities of fermionic and bosonic components coming from the superfield $Y_1$, so that the equilibrium value $n_Y^{EQ}$ is given by
\begin{eqnarray}
n_Y^{EQ} = 2 \int \frac{d^3p}{(2\pi)^3}
\left(\frac{1}{e^{\beta E_p} - 1} + \frac{1}{e^{\beta E_p} + 1}\right),
\end{eqnarray}
with $E_p$ being $E_p = (|\vec{p}|^2 + M_1^2)^{1/2}$. On the other hand, $n_X$ is the dark matter asymmetry produced by the decay of $Y$. The Hubble parameter is denoted by $H$, while $\Gamma_D$ stands for the thermally averaged decay width of $Y$ which is given by $\Gamma_D \simeq |\kappa_i|^2/(16\pi)$ in the non-relativistic limit. In our scenario, the asymmetry parameter $\epsilon$ is given by
\begin{equation}
\epsilon
=
-\frac{\Gamma_D(Y_1 \to \bar{X}^2) - \Gamma_D(Y_1 \to \bar{X}^{*2})}{\Gamma_D}
\simeq
-\frac{3}{4\pi} \frac{M_1}{M_2}
\frac{{\rm Im}[\kappa_1^2\kappa_2^{*2}]}{|\kappa_1|^2}.
\label{epsilon}
\end{equation}
The washout terms involved in the second Boltzmann equation are composed of two processes; the off-shell exchanges of $Y_i$ between $\bar{X}$ scatterings (2 $\leftrightarrow$ 2 washout effect) and the inverse decay of $Y_1$ (2 $\to$ 1 washout effect). First washout effect can be neglected compared to the second one when we consider the narrow-width region of the $Y_1$ decay, namely, $\Gamma_D \ll M_1$. The condition is easily satisfied when $|\kappa_1| \ll 1$. Furthermore, the second washout effect can also be neglected when we consider the weak washout regime, namely, $\Gamma_1 \ll H_1$, where $H_1$ is the Hubble parameter at $T = M_1$. The last condition is leading to the one on the $\kappa_1$ parameter to be
\begin{eqnarray}
|\kappa_1|^2
\ll \frac{32\pi^2}{3}\sqrt{\frac{\pi}{5}g_{*S}} \frac{M_1}{M_{\rm pl}}
\simeq 1.3 \times 10^{-10} \left(\frac{M_1}{10^6~{\rm GeV}}\right),
\end{eqnarray}
where $M_{\rm pl} \simeq 10^{19}$ GeV is the Plank mass and $g_{*S} = 232.5$ is the massless degrees of freedom at $T = M_1$. It can be seen that $\kappa_1$ should be suppressed to be ${\cal O}(10^{-5})$.

In the weak washout regime, the above Boltzmann equations lead to that
 the symptotic 
 asymmetry ($Asym.$) produced by the decay of $Y_1$ is simply given by
\begin{eqnarray}
Asym.
\equiv n_X(\infty)/s(\infty) = \epsilon \times n_Y^{EQ}(M_1)/s(M_1)
\simeq 5.2 \times 10^{-4} \epsilon,
\label{asymmetry}
\end{eqnarray}
where $s(T) = (2\pi^2/45)g_{*S}T^3$ is the entropy of the universe at the temperature $T$. Since the dark sector is not received severe constrain from experiments and thus the coupling constant $\kappa_2$ can be freely taken\footnote{The value of $\kappa_2$ should be less than {\cal O}(1) in order to verify the perturbative treatment of the calculations we have performed so far. Even such a case, enough asymmetry can be produced.}, we can easily generate enough asymmetry as can be seen in Eq.(\ref{epsilon}). Once the asymmetry $Asym.$ is generated, it is distributed to dark matter and $B-L$ asymmetries of the SM sector through the non-renormalizable operator $(\bar{X}^2LH_u)/\Lambda$ (ADM messenger operator) and the Spharelon process.

When the temperature of the universe is so high that both the $(\bar{X}^2LH_u)/\Lambda$ process and the Spharelon process are efficient, dark matter and anti-dark matter are in chemical equilibrium with SM particles which carry lepton (baryon) number. Note that the asymmetry $(B-L-DM/2)$ is preserved in this era. The dark matter asymmetry is related to baryon and lepton number asymmetries through the relation~\cite{Kaplan:2009ag},
\begin{equation}
B - L = -{79 \over 11} DM = -\frac{79}{169} (Asym.).
\label{relation 1}
\end{equation}
The baryon asymmetry is thus simply obtained through the Darkgenesis.

When the temperature of the universe becomes lower and the
 $(\bar{X}^2LH_u)/\Lambda$ process is frozen out (the Spharelon process
 is still active), both asymmetries $DM$ and $B-L$ are conserved
 individually. The dark matter asymmetry is therefore not altered after
 this era. Finally, when the temperature of the universe becomes as low
 as 100 GeV, the Spharelon process is also frozen out. After that, not
 only $DM$ but also $B$ and $L$ are conserved individually. The relation
 between $B$ and $B-L$ after taking account of finite mass effects is
 given by $B/(B-L)\simeq 0.31$~\cite{Kaplan:2009ag}. With the use of the
 relation in Eqs.(\ref{asymmetry}) and
 (\ref{relation 1}), the baryon asymmetry today turns out to be
\begin{eqnarray}
B
\simeq 0.31 \times \left[-\frac{79}{169} (Asym.)\right]
\simeq 7.5 \times 10^{-5} \epsilon.
\end{eqnarray}
Since the observed baryon asymmetry is $B^{(\rm obs)} \simeq 8.8 \times 10^{-11}$, the asymmetry parameter $\epsilon$ of order $10^{-6}$ is required, which is easily obtained by choosing an appropriate value of $\kappa_2$ in Eq.(\ref{epsilon}) unless $Y_2$ is extremely heavy compared to $Y_1$.

Finally, we consider the dark matter density which is also originally
produced by the decay of $Y_1$. Once the $(\bar{X}^2LH_u)/\Lambda$
process is frozen out, the dark matter asymmetry $DM$ is preserved. The
annihilation between $X$ and $\bar{X}$, however, is still active. In the NMSSM, it is possible to obtain very light scalar boson, which is composed dominantly of the scalar component of the $S$ field. In such a case, $X$ and $\bar{X}$ annihilates into two scalars by exchanging the singlino, which is the fermionic component of $S$, with the cross section larger than 1pb even if the dark matter is light. Due to this annihilation process, dark matter particle $X$ is annihilated away, and only anti-dark matter survives until today. Using the observed values of the baryon asymmetry $\Omega_B$ as well as the dark matter density $\Omega_{DM}$, the mass of dark matter is then estimated to be $M_{DM}\simeq 11$ GeV.\\

\noindent
{\large{\bf Conclusions}}\\

\vspace{-0.5cm}

\noindent
We have proposed a novel mechanism for generating the suitable baryon
asymmetry through a dark sector. This is a Baryogenesis through a
reverse pathway of the ADM scenario, where the dark matter asymmetry is
generated at first and then transferred into the suitable baryon
asymmetry in the SM sector. As in the case of the original ADM scenario,
the mechanism is possible to explain not only the baryon asymmetry but
also the dark matter density of the present universe. Since the origins
of these observables are coming from a dark sector which is not severely
constrained, it is possible to construct a concrete model with a very simple setup.\\

\noindent
{\large \bf Acknowledgments}\\

\vspace{-0.5cm}

\noindent
We thank K. Hamaguchi and T. Shindou for useful and helpful discussions. This work is partially supported by Scientific Grant by Ministry of Education and Science, Nos. 20540272, 20039006, 20025004, 21740174, and 22244021. We also thank the Yukawa Institute for Theoretical Physics at Kyoto University. Discussions during the YITP workshop ``Supper Institute 2010'' (YITP-W-10-07) were useful to complete this work.


\end{document}